# MODIFYING AND OPTIMIZING THE INVERSE OF THE FREQUENCY RESPONSE CIRCULANT MATRIX AS AN ITERATIVE LEARNING CONTROL COMPENSATOR

Shuo Liu[1] and Richard W. Longman[2]

Feedback control systems do not do what you ask. The concept of bandwidth is defined to tell what components of a command are reasonably well handled. Iterative Learning Control (ILC) seeks to converge to zero error following any given finite time desired trajectory as iterations progress. The approach can be used to achieve high precision tracking in spacecraft sensors performing repeated highly accurate sensor scanning. ILC asks for zero error for a finite time tracking maneuver, containing initial transients each iteration. The purpose of this paper is to create a method of designing ILC compensators based on steady state frequency response, and have the ILC converge to zero error in spite of transients and bandwidth. In this work the inverse of the circulant matrix of Markov parameters is used as a learning gain matrix. One can show that this matrix gives the steady state frequency response of the system at the finite number of frequencies observable in the finite data sequence of an iteration or run. Methods are used to adjust the steady state frequency response gains to address the transient part of the error signal. Numerical simulations compare the design approach to common time domain ILC design approaches, and one observes much faster convergence.

## INTRODUCTION

Feedback control systems do not do what you ask them to do. Concerning commands to a feedback control system, the concept of bandwidth is created to characterize what frequency components of a command will be reasonably well executed. Performance is similarly imperfect when the objective of the feedback control systems is to eliminate the influence of disturbances. When the feedback control system is to repeatedly perform the same task, or the command /disturbance is a periodic function of time, Iterative Learning Control (ILC) and Repetitive Control (RC) offer the possibility of addressing these standard feedback control difficulties, and in theory achieve zero tracking error for all frequencies up to Nyquist [1-4].

Spacecraft applications of RC include creating active vibration isolation mounts with the objective of eliminating the influence on fine pointing equipment of slight imbalance in reaction wheels or control moment gyros [5,6]. Reference [5] reports testing using a Stewart platform for isolation. Reference [6] uses a full floating spacecraft testbed with CMG's running, for laser communication (LaserCom). RC methods can use frequency response approaches for control design, because they have time progressing indefinitely, which means that one can wait till after initial condition transients have decayed to reach good performance. ILC applies to situations where a desired tracking maneuver is executed repeatedly, each time starting from the same initial condition, e.g., repeated scanning with a high precision sensor. Each run is a finite time run and contains an initial transient period.

[1] Department of Mechanical Engineering, Columbia University, 500 West 120th Street, New York, NY 10027
[2] Professor of Mechanical Engineering, Department of Mechanical Engineering, Columbia University, MC4703, 500 West 120th Street, New York, NY 10027

Feedback control design has two main approaches, time domain, and frequency domain. Frequency domain means design based on steady state frequency response. It is natural for RC since performance can wait for steady state after transients have decayed. ILC design is normally in the time domain because ILC asks to iteratively adjust the command to each run aiming to eventually achieve zero error throughout the finite time tracking objective -- including the initial transients phase.

Reference [7] examines one way to make use of the inverse of the frequency response apply to ILC. It makes use of a real time non-causal Finite Impulse Response (FIR) filter applied to the previous period data for RC design. The FIR gains are used to fill a matrix whose inverse is the learning gain matrix used each iteration to modify the command. There is a difficulty that at the beginning time steps and near the end time steps, the FIR filter terms need to be truncated since they ask for data before and after the finite time interval of the ILC problem. A method was used to modify a number of entries in the ILC learning control matrix to accomplish this, and to cause the inverse steady state frequency response to produce a convergent iterative sequence. This overcomes the apparent mismatch in the use of steady state frequency response thinking making it apply to the finite time ILC problem.

This paper examines another technique to accomplish all the same objectives, making steady state frequency response apply to designing ILC control laws. Use is made of the circulant form of the Toeplitz matrix of system Markov parameters [8]. In that reference the concept was motivated by designing frequency cutoff filters for robustification of ILC. In this paper we use the inverse of this circulant matrix of the system to create the ILC learning gain matrix. This matrix can be shown to directly produce the frequency response of the system for each frequency that one can observe in the number of time steps in the ILC problem. It produces the steady state response but based on a finite number of time steps, provided the number of steps at least goes beyond the settling time of the system. Again, we will need to adjust some gains, but this time the modifications are limited to addressing the mismatch, and there is no issue of having gains truncated. Designing ILC laws based on frequency response is attractive for a number of reasons, one of which is that one can make experimental frequency response tests of a system and then use the data directly without having to create a difference equation that tries to fit such data [9].

The ILC problem has another issue to contend with. ILC must be discrete time because it must use stored memory of error from the previous run or iteration executing the task. This means that design must be digital, and as a result somewhere a continuous time differential equations fed by a zero order hold is replaced by a difference equation whose solution is identical at the sample times. If the original differential equation has a pole excess of 3 or more, the discretization process will introduce a zero outside the unit circle [10] for reasonable sample rates. The ILC problem is an inverse problem, aiming to find that input that produces the desired output. We can say it is an ill-posed problem for such pole excesses, because inverse of the transfer function makes a zero outside the unit circle into an unstable pole. This is observed in the Toeplitz matrix of Markov parameters that serves as the time domain input output model for each run, by having a singular value and related singular vectors that are associated with the zero outside the unit circle [11]. Here we follow Reference [12] to create a well posed inverse problem by not asking for zero error at the first time step(s) of the finite time desired trajectory, the number of steps being equal to the number of zeros outside the unit circle.

The performance of ILC designs created by the methods presented here is compared in numerical examples to the performance of well known and well behaved ILC laws designed in the time domain [13-16]. Much faster convergence to zero tracking error is observed.

**GENERAL MATHEMATICAL FORMULATION OF ILC AND CONVERGENCE CONDITIONS**

**Basic Formulation of ILC**

This section reviews the general formulation of ILC as developed in [17]. Consider a SISO linear discrete-time control system in state space form:

$$x(k + 1) = Ax(k) + Bu(k) \quad k = 0,1,2, \dots, N - 1 \tag{1}$$
$$y(k) = Cx(k) \quad k = 1,2, \dots, N$$

Denote the desired $N$ time step output as $\underline{y^*}$, the actual output history at iteration $j$ as $\underline{y_j}$, the resulting tracking error history at iteration $j$ as $\underline{e_j} = \underline{y^*} - \underline{y_j}$, and the corresponding input history that produces the output at iteration $j$ as $\underline{u_j}$. Under bar indicates a column vector of the history of the variable for the iteration

$$\underline{y^*} = [y^*(1), y^*(2), \dots, y^*(N)]^T$$
$$\underline{y_j} = [y_j(1), y_j(2), \dots, y_j(N)]^T \tag{2}$$
$$\underline{e_j} = [e_j(1), e_j(2), \dots, e_j(N)]^T$$
$$\underline{u_j} = [u_j(0), u_j(1), \dots, u_j(N-1)]^T$$

Note that there is one time-step delay from input history to output history, corresponding to a zero order hold input-output delay. Simple modifications can be made for a feedback control system with two time steps delay, one for the controller, and one for the plant fed by a zero order hold. The general form of a linear ILC law updates the input history used in the previous iteration by a gain matrix $L$ times the error history observed in the previous run

$$\underline{u_{j+1}} = \underline{u_j} + L\underline{e_j} \tag{3}$$

Learning gain matrix $L$ has $N \times N$ learning gains chosen by the designer. The output history can be created by recursively applying Eq. (1), finding the convolution sum solution for each time step of an iteration

$$y_j(k) = CA^k x(0) + \sum_{i=0}^{N-1} CA^{k-i-1} B u_j(i) \tag{4}$$

and packaging the result in matrix form

$$\underline{y_j} = P\underline{u_j} + Ox(0) \tag{5}$$

Matrix $P$ is a lower triangular Toeplitz matrix containing the Markov parameters or the unit pulse response of the system. And $O$ is an $N \times 1$ step observability column vector

$$P = \begin{bmatrix} CB & 0 & \cdots & 0 \\ CAB & CB & \cdots & \vdots \\ \vdots & \vdots & \ddots & 0 \\ CA^{N-1}B & CA^{N-2}B & \cdots & CB \end{bmatrix} \tag{6}$$
$$O = [CA, CA^2, \dots, CA^N]^T$$

The ILC problem assumes that the system is reset to the same initial condition before each iteration. Based on Eq. (3) and (6), one can compute the error propagation matrix from one iteration to the next, and error in the first run to current error in the $j$th run

$$\underline{e_j} = (I - PL)\underline{e_{j-1}} = (I - PL)^j \underline{e_0} \tag{7}$$

**The If and Only If Condition for ILC Convergence to Zero Tracking Error and a Monotonic Decay Condition**

Equation (7) establishes that error $\underline{e_j}$ will converge to a zero vector as $j$ tends to infinity, for all initial run error histories $\underline{e_0}$, if and only if the spectral radius of the error propagation matrix $(I - PL)$ is less than one, i.e. the absolute values of all eigenvalues of this matrix are less than one

$$|\lambda_i(I - PL)| < 1 \tag{8}$$

There can be bad transients in the learning process. Making the singular value decomposition of the same matrix, and asking that all singular values be less than unity creates a condition for monotonic error decay from run to run in the sense of the Euclidean norm. Compute the singular value decomposition $I - PL = U_1 S_1 V_1^T$. Write the error history vector multiplied by input and output singular vector matrices, $U_1^T e_j = S_1 V_1^T e_{j-1}$. Since these matrices are unitary, the products do not change the Euclidean norm of the error. Then convergence is guaranteed to be monotonic if all singular values $\sigma_i$ on the diagonal of the singular value matrix $S_1$ are less than unity

$$\|e_j\| < \max(\sigma_i)\|e_{j-1}\| \tag{9}$$

This is a sufficient condition for convergence of the error history vector to the zero vector as $j$ tends to infinity, and a necessary and sufficient condition for monotonic convergence of the Euclidean error norm for all possible initial error histories in the initial run

$$\sigma_i(I - PL) < 1 \tag{10}$$

## THE CIRCULANT MATRIX OF MARKOV PARAMETERS OF A DISCRETE SYSTEM REFLECTS STEADY STATE FREQUENCY RESPONSE OF THE SYSTEM

A $P$ matrix can be generated from Eq. (6) by transferring a specific dynamic system from continuous-time system to discrete-time system with a zero order hold at a specific sample rate. A circulant matrix $P_c$ is a special kind of Toeplitz matrix in which the first column consists of Markov parameters and other columns follow a pattern that every entry of a selected column moves down one place, copying the final entry of the left neighboring column to the first vacated entry in the selected column

$$P_c = \begin{bmatrix} CB & CA^{N-1}B & \cdots & CA^2B & CAB \\ CAB & CB & \cdots & CA^3B & CA^2B \\ \vdots & \vdots & \ddots & \vdots & \vdots \\ CA^{N-2}B & CA^{N-3}B & \cdots & CB & CA^{N-1}B \\ CA^{N-1}B & CA^{N-2}B & \cdots & CAB & CB \end{bmatrix} \tag{11}$$

In terms of SISO system, the entries $CA^r B$ of the first column of matrix $P_c$ where $r = 0,1,2 \ldots$ work as the coefficients of a linear combination of several basic circulant matrices:

$$P_c = (CB)R_0 + (CAB)R_1 + \cdots + (CA^r B)R_r + \cdots + (CA^{N-1}B)R_{N-1} \tag{12}$$

where

$$(R_r)_{i,j} = \begin{cases} 1, & \text{for } i - j = r \text{ or } r - p \\ 0, & \text{otherwise} \end{cases} \tag{13}$$

In Eq. (13), every $R_r$ is an $N$ by $N$ matrix with only ones and zeros located in similar circluant format. Define $z_0 = e^{iw_0} = e^{\frac{2\pi i}{N}}$, $\underline{U} = U\left(e^{\frac{2\pi i}{N}f}\right) = [U_0\ U_1\ \cdots\ U_{N-1}]^T$, $\underline{Y} = Y\left(e^{\frac{2\pi i}{N}f}\right) = [Y_0\ Y_1\ \cdots\ Y_{N-1}]^T$, $\underline{u} = u(k)$ and $\underline{y} = y(k)$. Our input and output signals are $N$ steps long, the discrete Fourier transform of input and output signals are

$$\underline{U} = H\underline{u}$$
$$\underline{Y} = H\underline{y} \tag{14}$$

where

$$H = \begin{bmatrix} (z_0^0)^0 & (z_0^0)^{-1} & \cdots & (z_0^0)^{-(N-1)} \\ (z_0^1)^0 & (z_0^1)^{-1} & \cdots & (z_0^1)^{-(N-1)} \\ \vdots & \vdots & \ddots & \vdots \\ (z_0^{N-1})^0 & (z_0^{N-1})^{-1} & \cdots & (z_0^{N-1})^{-(N-1)} \end{bmatrix} \quad (15)$$

Consider $\underline{y} = P_c \underline{u}$ and $\underline{Y} = P_E \underline{U}$. Based on Eqs. (14) and (15)

$$P_E = HP_cH^{-1} = (CB)HR_0H^{-1} + \cdots + (CA^rB)HR_rH^{-1} + \cdots + (CA^{N-1}B)HR_{N-1}H^{-1} \quad (16)$$

and

$$HR_rH^{-1} = diag(1, z_o^{-r}, z_o^{-2r}, \cdots, z_o^{-(N-1)r}) \quad (17)$$

The $(i,j)$ component of matrix $P_E$ is generated by the sum of the corresponding entry

$$P_E(i,j) = \begin{cases} C(z_0^{j-1}I - A)^{-1}(I - A^{N-1}z_0^{-(j-1)(N-1)})B, & \text{for } i = j \\ 0, & \text{otherwise} \end{cases} \quad (18)$$

We assume that the number of time steps in each run is long enough that it reaches or goes beyond the time steps in the setting time of the system. Mathematically, this asks that $N$ be sufficiently large that matrix $A^{N-1}$ is negligible. Then the diagonal elements of circulant matrix $P_E$ are very close to $C(zI - A)^{-1}B$ where $z = z_0^{j-1}$. This is the same as the input-output transfer function matrix at frequency $j - 1$ from state space equations of the discrete SISO system if the initial states are zero. This means the circulant matrix $P_c$ of Markov parameters of a discrete system reflects the steady state frequency response corresponding to the $z$-transfer function analyzed at the integer discrete frequencies that can be seen in $N$ time steps. The same result can also be achieved for MIMO systems by computing the diagonal components of the corresponding matrix $P_E$.

**THE PERFORMANCE OF ILC USING UNALTERED INVERSE CIRCULANT MATRIX AS THE LEARNING GAIN MATRIX**

Consider a third order system

$$G(s) = \left(\frac{a}{s+a}\right)\left(\frac{\omega_0^2}{s^2 + 2\xi\omega_0 s + \omega_0^2}\right) \quad (19)$$

$$a = 8.8; \ \omega_0 = 37; \ \xi = 0.5$$

Reference [10] shows that when a continuous time system is fed by a zero order hold, zeros are introduced in the discretization process that are outside the unit circle, for sufficiently fast sample rate, and pole excesses of three or more. Reference [11] shows this is reflected in the singular value decomposition of the Toeplitz matrix $P$ in Eq. (6), exhibiting particularly small singular values of which the number matches the number of unstable zeros of the corresponding sampled system (for reasonable sample rates). ILC aims for zero error, and hence aims for use of the unstable inverse of matrix $P$. Here we use the inverse of the frequency response of the system, $P_c^{-1}$, which has the similar function as matrix $P^{-1}$. Create the learning gain matrix from $P_c^{-1}$ which does not have the instability of the inverse transfer function that is contained in $P^{-1}$. Compared to the $P^{-1}$, the singular values of $P_c^{-1}$ are much smaller and the first six singular values and eigenvalues of $(I - PP_c^{-1})$ are shown in Table 1 for matrix size $51 \times 51$.

**Table 1 First Six Singular Values and Eigenvalues of $(I - PP_c^{-1})$ from Large to Small**

| Order | $\sigma_1$ | $\sigma_2$ | $\sigma_3$ | $\sigma_4$ | $\sigma_5$ | $\sigma_6$ |
|---|---|---|---|---|---|---|
| Singular value | 18.2151 | 1.3772 | 0.2477 | 0.0034 | 0.0034 | 0.0033 |

| Order | $\lambda_1$ | $\lambda_2$ | $\lambda_3$ | $\lambda_4$ | $\lambda_5$ | $\lambda_6$ |
|---|---|---|---|---|---|---|
| Eigenvalue | 1 | 0.0033 | 0.0033 | 0.0033 | 0.0031 | 0.0031 |

Judged by the similar small eigenvalues of $P$ and $P_c$, the result of decomposition of the $P$ matrix is similar to that of $P_c$ matrix. Note that for the third order system from Eq. (19), the pole excess is 3, which gives the sampled system one unstable zero outside the unit circle provided the sampling period is sufficiently small, 0.02 second used here. This unstable zero makes the last singular value of $P$ matrix extremely small, but the $P_c$ matrix is not affected by the unstable zero. The decomposition of $(I - PP_c^{-1})$ is:

$$I - PP_c^{-1} = UIU^{-1} - U diag(\sigma_i) V^T (V_c^T)^{-1} diag(\sigma_{ci}^{-1}) U_c^{-1}$$

$$\approx UIU^{-1} - U diag(\sigma_i) V^T (V^T)^{-1} diag(\sigma_{ci}^{-1}) U^{-1}$$

$$= U \left( \begin{bmatrix} 1 & 0 & \cdots & 0 \\ 0 & 1 & \ddots & \vdots \\ \vdots & \ddots & \ddots & 0 \\ 0 & \cdots & 0 & 1 \end{bmatrix} - \begin{bmatrix} \sigma_1 & 0 & \cdots & 0 \\ 0 & \ddots & \ddots & \vdots \\ \vdots & \ddots & \sigma_{N-1} & 0 \\ 0 & \cdots & 0 & 0 \end{bmatrix} \begin{bmatrix} \sigma_{c1}^{-1} & 0 & \cdots & 0 \\ 0 & \sigma_{c2}^{-1} & \ddots & \vdots \\ \vdots & \ddots & \ddots & 0 \\ 0 & \cdots & 0 & \sigma_{cN}^{-1} \end{bmatrix} \right) U^{-1}$$

$$= U \begin{bmatrix} \varepsilon_1 & 0 & \cdots & 0 \\ 0 & \varepsilon_2 & \ddots & \vdots \\ \vdots & \ddots & \ddots & 0 \\ 0 & \cdots & 0 & 1 \end{bmatrix} U^{-1} \qquad (20)$$

where each $\varepsilon_i$ is very small. If the $U$ matrix is sorted into a specific arrangement $Q$ following the descending order of middle diagonal matrix, the eigenvalue decomposition of $(I - PP_c^{-1})$ is approximately

$$I - PP_c^{-1} \approx Q diag(\lambda_i) Q^{-1} = Q \begin{bmatrix} 1 & 0 & \cdots & 0 \\ 0 & \varepsilon_{N-1} & \ddots & \vdots \\ \vdots & \ddots & \ddots & 0 \\ 0 & \cdots & 0 & \varepsilon_1 \end{bmatrix} Q^{-1} \qquad (21)$$

which nearly tallies with eigenvalues observed in Table 1. The multiplication of two $(I - PP_c^{-1})$ is approximately equal to the single one but more closer to zero matrix than single except the unit spectral radius for the second to the last singular values are smaller, closer to zero

$$(I - PP_c^{-1})(I - PP_c^{-1}) \approx Q \begin{bmatrix} 1^2 & 0 & \cdots & 0 \\ 0 & \varepsilon_{N-1}^2 & \ddots & \vdots \\ \vdots & \ddots & \ddots & 0 \\ 0 & \cdots & 0 & \varepsilon_1^2 \end{bmatrix} Q^{-1} \approx I - PP_c^{-1} \qquad (22)$$

The next two experiments are achieved by analyzing the decomposition of $(I - PP_c^{-1})^3$ and $(I - PP_c^{-1})^6$ and the results are shown in Table 2 and 3. The values in Table 3 are smaller and better than those in Table 1 and 2, meaning the multiplication of inverse circulant matrix based error propagation matrix improves the computational accuracy to accelerate the process of $P_c^{-1}$ reaching $P^{-1}$. A better learning controller $L_{new}$ can also be obtained from this acceleration method in Eq. (23)

**Table 2 First Six Singular Values and Eigenvalues of $(I - PP_c^{-1})^3$ from Large to Small**

| Order | $\sigma_1$ | $\sigma_2$ | $\sigma_3$ | $\sigma_4$ | $\sigma_5$ | $\sigma_6$ |
|---|---|---|---|---|---|---|
| Singular value | 12.7055 | $1.1210e^{-5}$ | $7.4452e^{-7}$ | $5.2204e^{-8}$ | $3.9861e^{-8}$ | $3.5913e^{-8}$ |

| Order | $\lambda_1$ | $\lambda_2$ | $\lambda_3$ | $\lambda_4$ | $\lambda_5$ | $\lambda_6$ |
|---|---|---|---|---|---|---|
| Eigenvalue | 1 | $3.5853e^{-8}$ | $3.4553e^{-8}$ | $3.4553e^{-8}$ | $3.0968e^{-8}$ | $3.0968e^{-8}$ |

Table 3 First Six Singular Values and Eigenvalues of $(I - PP_c^{-1})^6$ from Large to Small

| Order | $\sigma_1$ | $\sigma_2$ | $\sigma_3$ | $\sigma_4$ | $\sigma_5$ | $\sigma_6$ |
|---|---|---|---|---|---|---|
| Singular value | 12.7055 | $2.6757e^{-13}$ | $1.3630e^{-14}$ | $1.2497e^{-15}$ | $1.1918e^{-15}$ | $1.1918e^{-15}$ |

| Order | $\lambda_1$ | $\lambda_2$ | $\lambda_3$ | $\lambda_4$ | $\lambda_5$ | $\lambda_6$ |
|---|---|---|---|---|---|---|
| Eigenvalue | 1 | $3.1086e^{-15}$ | $1.5776e^{-15}$ | $8.9522e^{-16}$ | $1.1039e^{-15}$ | $1.1039e^{-15}$ |

$$L_{new} = P^{-1}[I - (I - PP_c^{-1})^6] \qquad (23)$$

Consider a special case that tracking error fails to decay monotonically with learning controller $L_{new}$. The similar phenomenon is also described in Reference [7]. Let the singular value decomposition of $(I - PL_{new})$ be $U_n S_n V_n^T$ with $\underline{u_i}$ and $\underline{v_i}$ the $i^{th}$ columns of $U_n$ and $V_n$ associated with singular value $\sigma_i$. The initial error vector that fails to learn monotonically is $\underline{e_0} = \underline{v_1}$. Figure 1 shows the RMS of error vector for iterative learning steps 0 to 10. Since all singular values of $(I - PL_{new})$ are essentially zero except $\sigma_1$, $(I - PL_{new}) \approx \sigma_1 \underline{u_1} \underline{v_1}^T$ and the error $\underline{e_1}$ is $\sigma_1 \underline{u_1}$ if $\underline{e_0} = \underline{v_1}$. From Eq. (22), $(I - PL_{new})(I - PL_{new})$ is almost $(I - PL_{new})$. The equation

$$\left(\sigma_1 \underline{u_1} \underline{v_1}^T\right)\left(\sigma_1 \underline{u_1} \underline{v_1}^T\right) = \sigma_1 \underline{u_1}\left(\underline{v_1}^T \sigma_1 \underline{u_1}\right)\underline{v_1}^T \approx \sigma_1 \underline{u_1} \underline{v_1}^T \qquad (24)$$

will be satisfied if and only if $\underline{v_1}^T \sigma_1 \underline{u_1} \approx 1$. The tracking errors starting from the first iteration will stay constant as shown by

$$\underline{e_j} = (I - PL_{new})\underline{e_{j-1}} = (I - PL_{new})^{j-1}\underline{e_1}$$
$$\approx \left(\sigma_1 \underline{u_1} \underline{v_1}^T\right)^{j-1} \sigma_1 \underline{u_1} = \sigma_1 \underline{u_1} \underline{v_1}^T \sigma_1 \underline{u_1} = \sigma_1 \underline{u_1} \qquad (25)$$

This reveals one drawback of the learning controller $L_{new}$ that the maximum singular value $\sigma_1$ bigger than one will destroy monotonic decay of tracking error for some parts of the input-output spaces, but the error sharply diminishes for the remaining components of the output signals. The unstable zero outside the unit circle of the sampled third order system throws the maximum singular value $\sigma_1$ outside of the unity boundary and brings the unit eigenvalue $\lambda_1$ together with the extremely small singular value $\sigma_N$ that cannot be eliminated from the non-deleted $P$ matrix. This phenomenon causes the inverse problem to have the input command needed for desired output grow extremely large [11]. To avoid the unstable inverse issue and to make the learning controller universal, we simply delete the first row of matrix $P$ and first column of matrix $P_c^{-1}$ so that we are no longer asking for zero error at one initial step, which makes $P, P_c^{-1}$ into $P_1, P_{c1}^{-1}$. The extremely small singular value disappears in terms of one-row-deleted $P$ matrix and it should be feasible to reduce the largest singular value and eigenvalue below one.

**INVESTIGATE TRACKING ERROR AT REMAINING TIME STEPS WHEN ELIMINATING THE LEARNING ACTION OF THE FIRST TIME STEP**

Now consider the 50 Hz sampled third order system from Eq. (19) with the original matrix size $51 \times 51$ and calculate the first six singular values of $(I - P_1 P_{c1}^{-1})$ as shown in Table 4.

Table 4 First Six Singular Values and Eigenvalues of $(I - P_1 P_{c1}^{-1})$ from Large to Small

| Order | $\sigma_1$ | $\sigma_2$ | $\sigma_3$ | $\sigma_4$ | $\sigma_5$ | $\sigma_6$ |
|---|---|---|---|---|---|---|
| Singular value | 13.8093 | 0.5417 | 0.1135 | 0.0034 | 0.0034 | 0.0033 |
| Order | $\lambda_1$ | $\lambda_2$ | $\lambda_3$ | $\lambda_4$ | $\lambda_5$ | $\lambda_6$ |
| Eigenvalue | 0.9987 | 0.0032 | 0.0032 | 0.0031 | 0.0031 | 0.0030 |

The first singular value exceeds the monotonic decaying boundary. Knowing that $\sigma_1 > \sigma_2$, $\sigma_1 \geq |\lambda_{max}|$, the tracking error norm can satisfy both monotonic decay and convergence to zero by reducing $\sigma_1$ lower than one.

**Adjust the Overall Gain to Improve the ILC Performance**

Based on Eq. (3), consider simply introducing an overall gain $\Phi$ in front of the deleted inverse circulant matrix $P_{c1}^{-1}$ to form the learning law

$$\underline{u_{j+1}} = \underline{u_j} + \Phi P_{c1}^{-1} \underline{e_j} \tag{26}$$

By changing the overall gain from -1 to 2, we notice in Figure 2 that the largest singular value of $(I - \Phi P_1 P_{c1}^{-1})$ reaches the smallest value 1 when $\Phi$ is 0, which means one cannot produce monotonic convergence without the need to adjust any gains inside the deleted inverse circulant matrix.

**Adjust Some Gains of the Deleted Inverse Circulant Matrix to Improve ILC Performance**

*What Gains to Adjust.* In this section we mainly focus on reducing the maximum singular value. First, we need to know what gains we should adjust to be most effective in decreasing $\sigma_1$. Calculate the sensitivity derivatives $\frac{\partial \sigma_1}{\partial l_{ij}}$ of the maximum singular value to each individual gain in matrix $P_c^{-1}$. In order to find the relationship between $H$ and $\sigma_1$, define $H = I - P_1 P_{c1}^{-1}$, then the singular value decomposition of $(I - P_1 P_{c1}^{-1})$ is $U_2 S_2 V_2^T$ where $S_2$ is $diag(\sigma_1, \sigma_2, ..., \sigma_{N-1})$, $U_2$ is matrix $[u_1, u_2, ..., u_{N-1}]$ and $V_2$ is matrix $[v_1, v_2, ..., v_{N-1}]$, which makes $H = \sigma_1 u_1 v_1^T + \sigma_2 u_2 v_2^T + \cdots + \sigma_{N-1} u_{N-1} v_{N-1}^T$. Note that every $u_i$ or $v_i$ is a unitary orthogonal vector, so the maximum singular value can be expressed as $\sigma_1 = u_1^T H v_1$. Using the result of derivatives of eigenvalues of symmetric matrices, the derivative of a singular value with respect to an entry in the $H$ matrix is

$$\frac{\partial \sigma_k}{\partial l_{ij}} = u_k^T \frac{\partial H}{\partial l_{ij}} v_k \tag{27}$$

Note that

$$\frac{\partial H}{\partial l_{ij}} = -P_1 \frac{\partial P_{c1}^{-1}}{\partial l_{ij}} \tag{28}$$

Therefore, we can compute the sensitivity of any singular value to any gain $l_{ij}$ from

$$\frac{\partial \sigma_k}{\partial l_{ij}} = -u_k^T P_1 \frac{\partial P_{c1}^{-1}}{\partial l_{ij}} v_k \tag{29}$$

The matrix $\frac{\partial P_{c1}^{-1}}{\partial l_{ij}}$ is a matrix of the same dimension as $P_{c1}^{-1}$ with all gains zero except for a unit gain in the same location as gain $l_{ij}$. We use this result to generate multidimensional Figures 3 and 4 describing what gains have the most influence on the change of maximum singular value $\sigma_1$. These two figures reveal that the gains from the first and last five columns have sharp or gentle fluctuation in sensitivity, hence we establish the steepest descent optimization algorithm to reduce the maximum singular value corresponding to those unsteady gains.

*How to Adjust Those Gains.* Put all gains $l_{ij}$ inside a column vector $l'_i$, then the steepest descent algorithm generalizes to $\sigma_1(l'_{i+1}) = \sigma_1(l'_i + \Delta l'_i) \approx \sigma_1(l'_i) + S_i^T \Delta l'_i$, where $S_i = \frac{\partial \sigma_1}{\partial l'_i}$ and

$$\Delta l'_i = -[S_i S_i^T + rI]^{-1} S_i \sigma_1(l'_i) \tag{30}$$

which results from minimizing the quadratic cost function:

$$J_{i+1} = [\sigma_1(l'_i + \Delta l'_i)]^T [\sigma_1(l'_i + \Delta l'_i)] + r[\Delta l'_i]^T [\Delta l'_i] \tag{31}$$

Reevaluate the sensitivity derivatives and weight factor with new singular input and output column vectors for iteration $i + 1$, place all addressed gains $l_{ij}$ inside a column vector $l'_{i+1}$ and update again until the last loop, then the optimized inverse circulant law becomes $L = P_{oc1}^{-1}$.

Adjusting gains from the 5 by 5 area of the upper left corner and that of the upper right 5 by 5 corner was found to be the most effective method. Note that we did not repeat the sensitivity and weight factor evaluation procedure to modify the gains chosen during the iterations. After 1000 iterations with weight factor $r$ as 0.1, the maximum singular value is reduced to 0.2224, which corresponds to very fast monotonic learning with iterations. The downtrend of largest singular value together with largest eigenvalue is shown in Figure 5.

## PERFORMANCE COMPARISON WITH COMMON DESIGN APPROACHES FOR TIME DOMAIN ITERATIVE LEARNING CONTROL DESIGNS

The original design of ILC learning law [1] did not consider the instability of the inverse problem in discrete-time systems for pole excess of 3 or more. Before learning to create a stable inverse [12], the guaranteed good performance of ILC was limited to pole excesses of 1 or 2. But often in practice, the instability of the inverse problem was slow enough that it was not observed in practice. The time domain ILC designs, can be converted to well posed problems with stable control actions by eliminating initial time step or steps from being part of the learning process. We consider the time domain ILC laws for comparison: the partial isometry law $L_2 = V_3 U_3^T$, the contraction mapping law $L_3 = P^T$, and the quadratic cost law $L_4 = (P^T P + I)^{-1} P^T$ where $P = U_3 S_3 V_3^T$ which all can ensure monotonic convergence of the Euclidean norm of the tracking error [13-16]. In next part we will compare the tracking error norm (RMS of tracking error) of $L_2, L_3$ and $L_4$ with our optimized inverse circulant law $L_1 = P_{oc}^{-1}$ by both deleting and not deleting the first row of $P$ and first column of $L_i$ with different desired output signals $y_{d1} = \pi \left(1 - \cos\frac{\pi}{2} t\right)^2$ and $y_{d2} = \pi(5t^3 - 7.5t^4 + 3t^5)$ at 50 Hz sample rate. Note the maximum singular value is 1.3017 through the same optimization method in the previous section without deleting first row of $P$ and first column of $P_{oc}^{-1}$.

These two desired output trajectories we refer to as smooth start up trajectories. The error during the initial time step or steps, is small because of the "smooth" start. By considering the continuous time ILC problem instead of the discrete time system, and considering that the system starts at rest at time zero, the desired trajectory and at least its first two derivatives need to be zero at the start. Otherwise, the continuous time inverse problem requires the control action to apply an impulse and derivative of an impulse at time zero to get onto the desired trajectory. This need is not as striking in discrete time but we still suggest that the ILC user pick commands that do not require generalized functions to perform in the continuous time version

The comparison in Figure 6 and 7 is made at each iteration testing the RMS of the tracking error associated with or without asking convergence of the first time-step error. In both cases the optimized inverse circulant law has far faster learning with iterations than any of the time domain ILC laws. It will be even faster if the acceleration method is used to generate the new learning controller $L_{new}$ as from Eq. (25).

## EFFECTIVENESS OF ADJUSTING GAINS IN HIGHER ORDER SYSTEMS

Consider a fourth order system

$$G(s) = \left(\frac{\omega_0^2}{s^2+2\xi\omega_0 s+\omega_0^2}\right)\left(\frac{\omega_1^2}{s^2+2\xi\omega_1 s+\omega_1^2}\right) \tag{32}$$

$$\omega_0 = 37; \ \omega_1 = 74; \ \xi = 0.5$$

and a fifth order system

$$G(s) = \left(\frac{a}{s+a}\right)\left(\frac{\omega_0^2}{s^2+2\xi\omega_0 s+\omega_0^2}\right)\left(\frac{\omega_1^2}{s^2+2\xi\omega_1 s+\omega_1^2}\right) \tag{33}$$

$$a = 8.8; \ \omega_0 = 37; \ \omega_1 = 74; \ \xi = 0.5$$

We use the same sensitivity derivatives method from Eq. (27) to see what gains in deleted inverse circulant matrix affect the singular value the most. The pole excess of the fourth and the fifth order systems are 4 and 5, with one and two zeros introduced outside the unit circle during discretization. We delete the first two rows of the corresponding matrix $P$ as $P_2$ and the first two columns of the corresponding inverse circulant matrix $P_c^{-1}$ as $P_{c2}^{-1}$. The sensitivity derivatives of these two systems in Figure 8 reveal that the gains from the upper right block or the upper left as well as the upper right blocks of $P_{c2}^{-1}$ have sharp or gentle fluctuation in sensitivity, here we just adjust the same gains chosen in third order system (5 by 5 block of upper left corner and 5 by 5 block of upper right corner) to test the effectiveness of the steepest descent optimization algorithm in higher order systems.

After 10000 iterations with weight factor 0.1, the maximum singular value is reduced to 1.2395, and the spectral radius to 0.0052 for fourth order system. For fifth order system, the optimized singular value is 8.7425 and the optimized spectral radius is 0.3436. Next we will compare the tracking error norm of $L_2, L_3$ and $L_4$ with our optimized inverse circulant law $L_1 = P_{oc}^{-1}$ by deleting the first two rows of $P$ and first two columns of $L_i$ with different desired output signals $y_{d1} = \pi\left(1 - \cos\frac{\pi}{2}t\right)^2$ and $y_{d2} = \pi(5t^3 - 7.5t^4 + 3t^5)$ at 50 Hz sample rate.

In Figures 9 and 10, the optimized inverse circulant law learns much faster than the other three learning laws. The shape of the RMS error curve mainly depends on the type of desired output signal and the singular values of the error propagation matrix. If the whole singular values are very small like the simulation results in the third, fourth and fifth order systems we tested, the learning speed of our designed learning law will be very fast even though the maximum singular value is above one. In this case, the special desired output signal such as the first column of the output singular matrix of the error propagation matrix will make the inverse circulant controller lose the function for learning, which means the RMS error keeps constant after the first iteration as the same in Figure 1.

## CONCLUSIONS

An ILC design process that uses the inverse circulant matrix of system Markov parameters as a learning gain matrix is shown to produce a particularly effective ILC law. This matrix is shown to represent the steady state frequency response of the system for finite time trajectories, provided the trajectory is long enough to contain the settling time. This means that the designs for the finite time trajectories of ILC are effectively being made based on steady state frequency response modelling. Thus, both the bandwidth limitations normally encountered in feedback control design, and the apparent mismatch of finite time ILC with steady state frequency response thinking is overcome. The design needs to address the issue of unstable inverse of most ILC problems by not asking for zero error at the first or first few time steps of the desired trajectory, an issue common to all ILC design. The proposed design method is shown to have particularly fast learning when compared to common ILC designs made in the time domain.

## REFERENCES

[1] S. Arimoto, S. Kawamura, and F. Miyazaki, "Bettering Operation of Robots by Learning," *Journal of Robotic. Systems*, Vol. 1, No. 2, 1984, pp. 123–140.

[2] Z. Bien and J.-X. Xu, editors, *Iterative Learning Control: Analysis, Design, Integration and Applications*, Kluwer Academic Publishers, Boston, 1998, pp. 107-146.

[3] H.-S. Ahn, Y. Chen, and K. L. Moore, "Iterative Learning Control: Brief Survey and Categorization," *IEEE Transactions on Systems, Man, and Cybernetics*, Part C, Vol. 37, No. 6, 2007, pp. 1099-1122.

[4] D. A. Bristow, M. Tharayil, A. G. Alleyne, "A Survey of Iterative Learning Control," *IEEE Control Systems Magazine*, Vol. 26, Issue 3, 2006, pp. 96-114.

[5] S. G. Edwards, B. N. Agrawal, M. Q. Phan, and R. W. Longman, "Disturbance Identification and Rejection Experiments on an Ultra Quiet Platform," *Advances in the Astronautical Sciences*, Vol. 103, 1999, pp. 633-651.

[6] E. S. Ahn, R. W. Longman, J. J. Kim, and B. N. Agrawal, "Evaluation of Five Control Algorithms for Addressing CMG Induced Jitter on a Spacecraft Testbed," *The Journal of the Astronautical Sciences*, Vol. 60, Issue 3, 2015, pp. 434-467.

[7] B. Panomruttanarug, R. W. Longman, and M. Q. Phan, "Steady State Frequency Response Design of Finite Iterative Learning Control," *The Journal of the Astronautical Sciences*, 2019, https://doi.org/10.1007/s40295-019-00198-9.

[8] B. Song and R. W. Longman, "Circulant Zero-Phase Low Pass Filter Design for Improved Robustification of Iterative Learning Control," *Advances in the Astronautical Sciences*, Vol. 156, 2016, pp. 2161-2180.

[9] B. Panomruttanarug and R. W. Longman, "Designing Optimized FIR Repetitive Controllers from Noisy Frequency Response Data," *Advances in the Astronautical Sciences*, Vol. 127, 2007, pp. 1723-1742.

[10] K. Åström, P. Hagander, and J. Strenby, "Zeros of Sampled Systems," *Proceedings of the 19th IEEE Conference on Decision and Control*, 1980, pp. 1077-1081.

[11] Y. Li and R. W. Longman, "Characterizing and Addressing the Instability of the Control Action in Iterative Learning Control," *Advances in the Astronautical Sciences*, Vol. 136, 2010, pp. 1967-1985.

[12] X. Ji, T. Li, and R. W. Longman, "Proof of Two Stable Inverses of Discrete Time Systems," *Advances in the Astronautical Sciences*, Vol. 162, 2018, pp. 123-136.

[13] J. Bao and R.W. Longman, "Unification and Robustification of Iterative Learning Control Laws," *Advances in the Astronautical Sciences*, Vol. 136, 2010, pp. 727–745.

[14] H. S. Jang and R. W. Longman, "A New Learning Control Law with Monotonic Decay of the Tracking Error Norm," *Proceedings of the Thirty-Second Annual Allerton Conference on Communication, Control, and Computing*, Monticello, IL, September 1994, pp. 314-323.

[15] H. S. Jang and R. W. Longman, "Design of Digital Learning Controllers Using a Partial Isometry," *Advances in the Astronautical Sciences*, Vol. 93, 1996, pp. 137-152.

[16] D.H. Owens and N. Amann, "Norm-Optimal Iterative Learning Control," *Internal Report Series of the Centre for Systems and Control Engineering*, University of Exeter, 1994.

[17] M. Phan and R. W. Longman, "A Mathematical Theory of Learning Control for Linear Discrete Multivariable Systems," *Proceedings of the AIAA/AAS Astrodynamics Conference*, Minneapolis, Minnesota, August 1988, pp. 740-746.

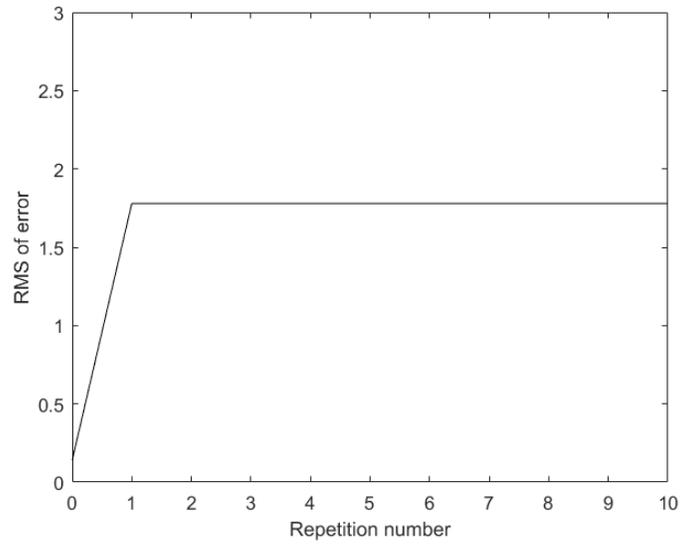

**Figure. 1 The RMS of error vs. iteration with initial error given by the first singular output vector $v_1$, input $u_0$ equal to 0 and desired output $y_d$ equal to $v_1$**

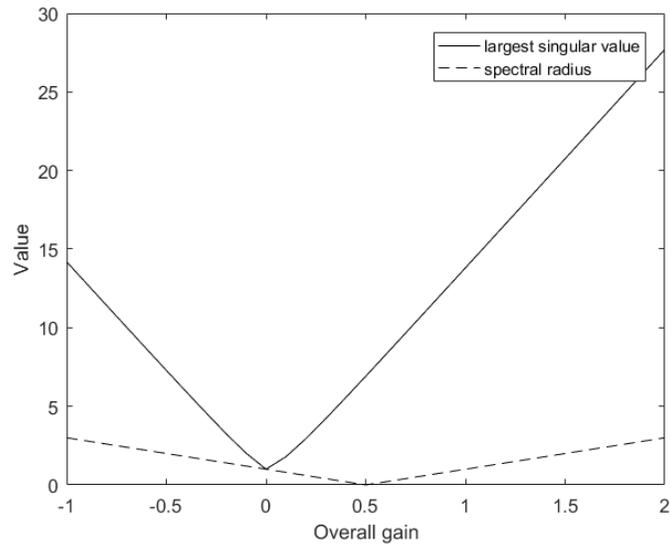

**Figure. 2 The largest singular value and the spectral radius as a function of changing the overall gain $\Phi$**

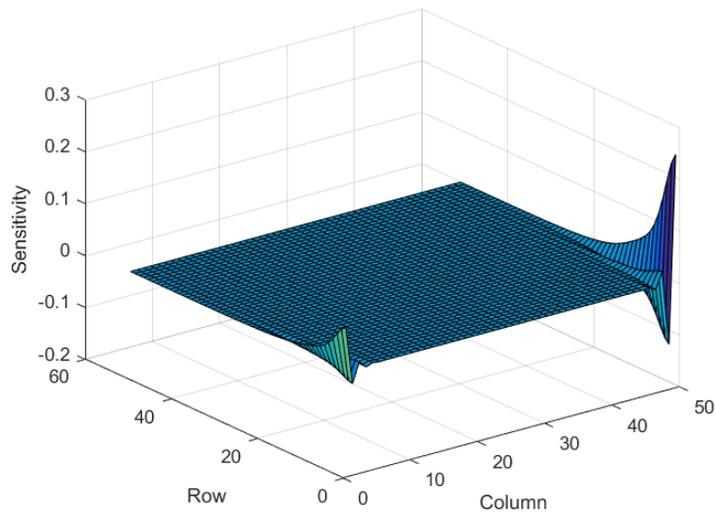

Figure. 3 Sensitivity of maximum singular value to all entries in $P_{c1}^{-1}$

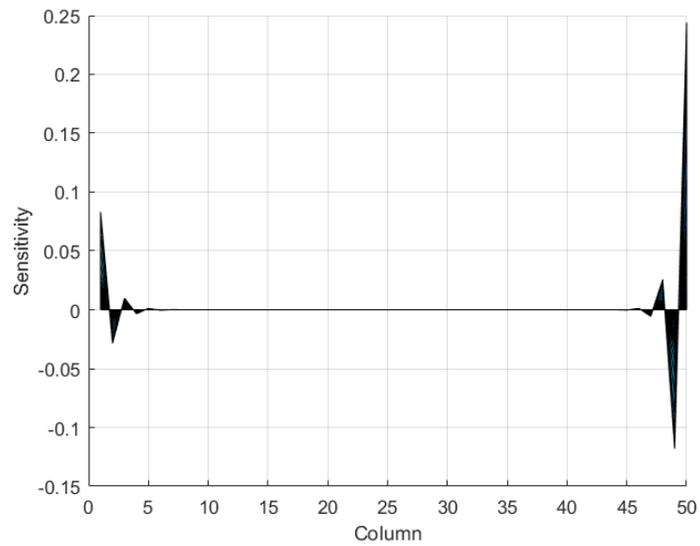

Figure. 4 Same as Figure 3 but in Row-Sensitivity plane

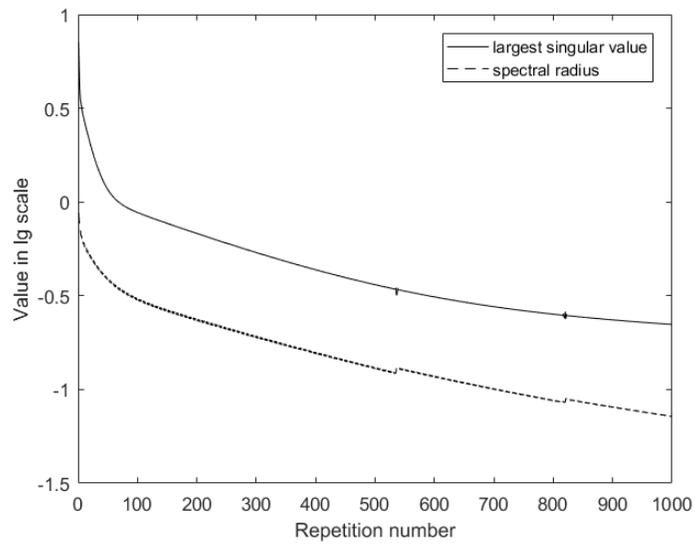

**Figure. 5 The largest singular value and spectral radius as a function of gain adjustment**

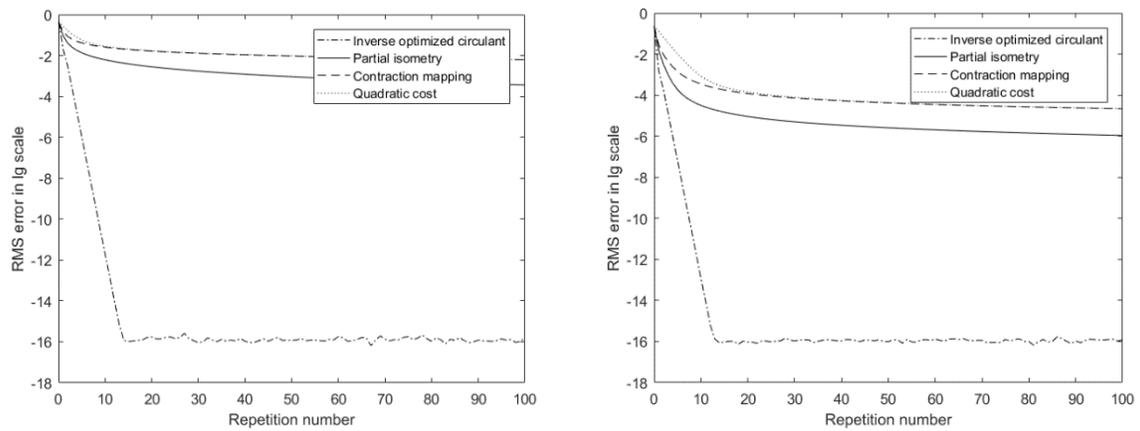

**Figure. 6 RMS error vs. iterations for deleted matrix with input $y_{d1}$ (left) and input $y_{d2}$ (right)**

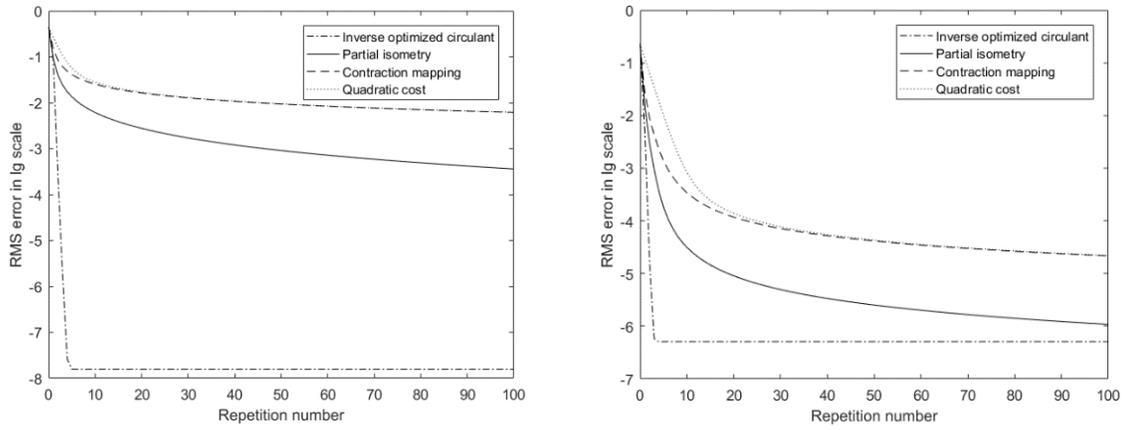

**Figure. 7 RMS error vs. iterations for non-deleted matrix with input $y_{d1}$ (left) and input $y_{d2}$ (right)**

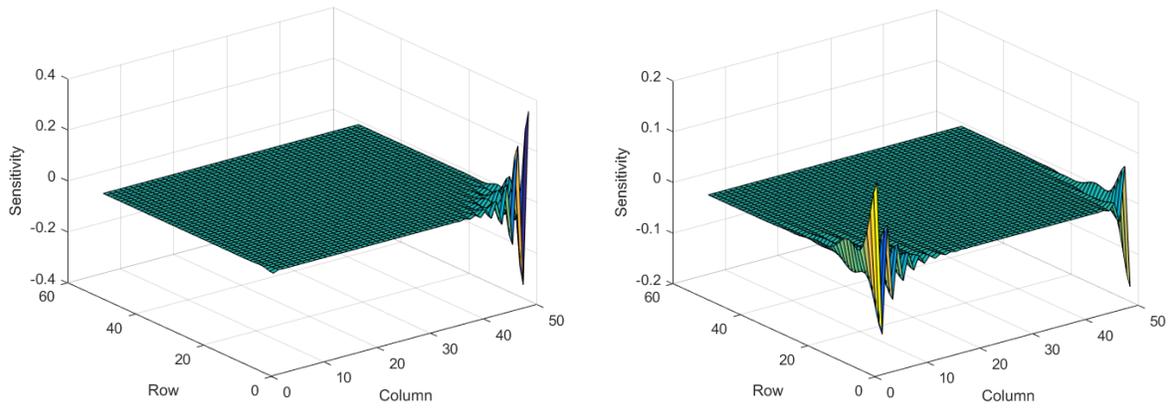

**Figure. 8 Sensitivity of maximum singular value to all entries in $P_{c2}^{-1}$ in 4$^{th}$ (left) and 5$^{th}$ (right) order systems**

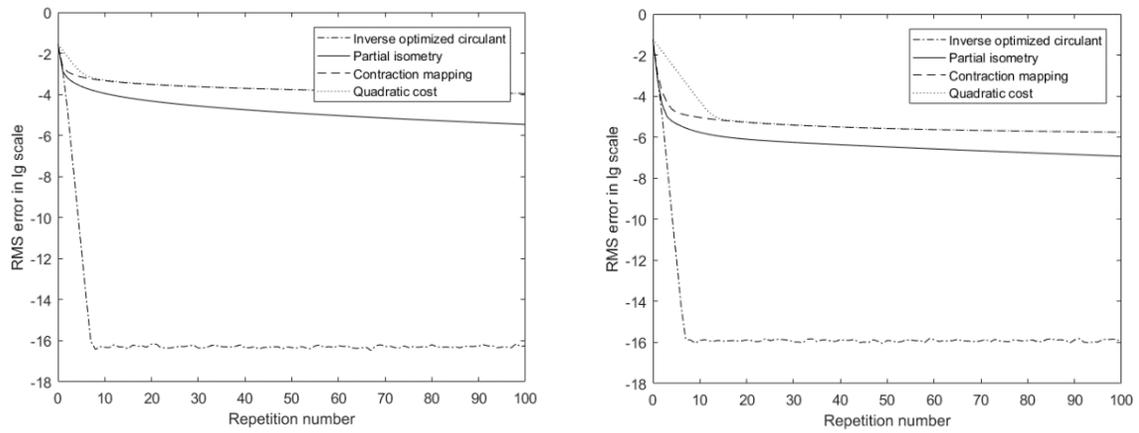

**Figure. 9 RMS error vs. iterations for deleted matrix with input $y_{d1}$ (left) and input $y_{d2}$ (right) in 4th order system**

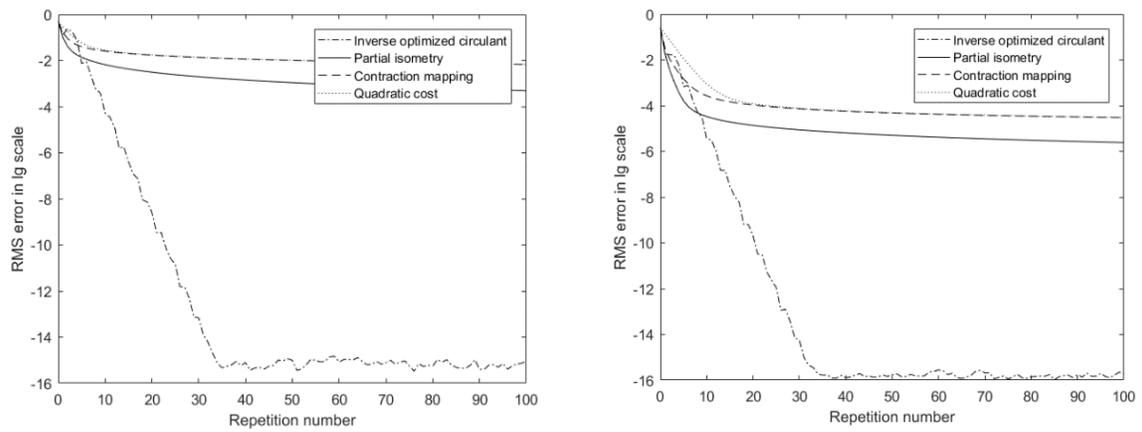

**Figure. 10 RMS error vs. iterations for deleted matrix with input $y_{d1}$ (left) and input $y_{d2}$ (right) in 5th order system**